# Molecular Recollision Interferometry in High Harmonic Generation

Xibin Zhou, Robynne Lock, Wen Li, Nick Wagner, Margaret. M. Murnane and Henry C. Kapteyn

*JILA and Department of Physics, University of Colorado and NIST, Boulder, CO 80309, USA*

*Ph. (303) 4927766; E-mail: Xibin.Zhou@colorado.edu*

**Abstract**

We use extreme-ultraviolet interferometry to measure the phase of high-order harmonic generation from transiently aligned $CO_2$ molecules. We unambiguously observe a reversal in phase of the high order harmonic emission for higher harmonic orders with a sufficient degree of alignment. This results from molecular-scale quantum interferences between the molecular electronic wave function and the recolliding electron as it recombines with the molecule, and is consistent with a two-center model. Furthermore, using the combined harmonic intensity and phase information, we extract accurate information on the dispersion relation of the returning electron wavepacket as a function of harmonic order. This analysis shows evidence of the effect of the molecular potential on the recolliding electron wave.

PACS numbers: 42.65Ky, 42.65Re

There has been considerable recent interest in using high harmonic generation (HHG) to probe the structure and dynamics of molecules [1-5]. In typical experiments, a pump pulse is used to excite a rotational wave packet in a molecular sample through an impulsive Raman process. At certain time delays after the pump, the excited wave packet will rephase to form a revival [6, 7]. A delayed strong laser pulse can then generate high-order harmonics whose intensity varies with the changing alignment distribution of the molecules. In the three-step model of HHG [8], the electric field of a focused laser ionizes an electron from the highest occupied molecular orbital (HOMO) of a molecule. This electron is then accelerated by the electric field of the laser. If it returns to the vicinity of the parent ion, a high-energy recollision event occurs, and the electron can recombine into the HOMO. This continuum-bound transition converts the kinetic energy of the electron into an EUV photon; information about the molecular structure is encoded in the orientation-dependence of the HHG intensity. The shape and symmetry of the HOMO can result in multi-center quantum interferences as the electron recombines. This is in exact analogy to the same effects long-recognized in the context of EUV photoelectron spectroscopy of molecules [9], since recombination in HHG is the process inverse to photoionization.

Lein et al. theoretically investigated quantum interferences in HHG from hydrogen molecules by solving the time dependent Schrödinger equation [10, 11]. A spectral minimum was found both in the harmonic spectrum for a fixed orientation, and in the angular dependence of the harmonic intensity for given harmonic order. Furthermore, this work predicted a sudden phase jump at the spectral minimum, where the phase of the electric field of the high harmonic emission reverses. These results can be understood as interference of harmonic emission from two spatially separated regions in a molecule.

In past work, Kanai et al. and Vozzi et al. attributed the observed alignment dependence of the harmonic emission from $CO_2$ molecules to quantum interference [3, 12]. However, the harmonic orders most strongly modulated by alignment differed in the two experiments (the $27^{th}$ vs the $33^{rd}$ orders, respectively). Furthermore, the authors used different relationships between the harmonic photon energy and the recolliding electron kinetic energy, $E_k = nh\omega$ or $E_k = nh\omega - I_p$, to explain the experimental data, where $nh\omega$ is the harmonic energy and $I_p$ is the molecular ionization potential. These differing relationships correspond to slightly different physical pictures -- the latter corresponds to the case where the recolliding electron interacts with the HOMO and generates emission when it returns to the edge of the molecular potential, while the former implies that HHG occurs when the electron returns to the "bottom" of the molecular potential, and is thus accelerated by the Coulomb field of the molecule ion. Other than interference, ground state depletion was also proposed as a possible mechanism for the observed modulations [13]. Therefore more accurate measurements of both the phase and intensity of HHG from aligned molecules are important, both to confirm conclusively that quantum interferences do exist in molecular HHG, and to provide insight on distortion of the recolliding wave function by the Coulomb potential of the ionized molecule. Measurement of the phase of the high harmonic emission is also critical for new approaches to molecular imaging [1]. While past experiments [14, 15] provided some evidence for phase jumps in HHG, they were inconclusive in making accurate comparisons with the two-center model.

In this work we present the first direct and accurate measurement of the phase of harmonic emission from molecules. We directly confirm that a two-center interference in the $CO_2$ HHG emission induces a minimum at certain angular distributions, with a corresponding phase shift of $\pi$. We fit the harmonic emission as a function of time within a rotational revival to the two-center interference model convolved with the rotational distribution, which allows us to extract the ratio of the two-center separation in the molecule to the returning electron wavelength as a function of harmonic order. The data show that the contribution from the molecular potential on the electron kinetic energy varies with harmonic order. These results have important implications for efforts to extract molecular structure from measurements of the harmonic emission from molecules, and can be used to benchmark realistic theories of molecules in strong fields.

$CO_2$ is an excellent candidate for studying harmonic phase effects because it is highly polarizable (easy to align), and the HOMO has a "two-center" character similar to $H_2$, but anti-symmetric rather than symmetric. It can be written approximately as a linear combination of two atomic $p$ orbitals $\left[\varphi_p(\mathbf{r}-\mathbf{R}/2)-\varphi_p(\mathbf{r}+\mathbf{R}/2)\right]/\sqrt{2}$, where $R$ is the distance between the two centers of electron density in the molecule. The recombination dipole thus has the form $D(\theta)=\langle e^{i\mathbf{k}\cdot\mathbf{r}}|\hat{O}|\Phi\rangle=-\sqrt{2}i\sin(kR\cos\theta/2)\int e^{-i\mathbf{k}\cdot\mathbf{r}}O\varphi(\mathbf{r})d\mathbf{r}$ [11], where $\theta$ is the angle between the molecular axis and the laser polarization. The wavelength of the returning electron wave can be calculated from $\lambda=\hbar/\sqrt{2mE_k}$ where $m$ is the electron mass and $E_k$ is the kinetic energy of the electron. The condition for the interference minimum is $R\cos\theta_c=\lambda$, which is a condition that is satisfied within the harmonic spectral range that is experimentally accessible. For a harmonic of wavelength $\lambda$, the sign of the harmonic emission dipole $D(\theta)$ reverses for angle below and above $\theta_c$. However, the harmonic emission from a molecular sample must be averaged over the angular distribution of the rotational wavepacket at the time of emission: $\int_0^\pi \rho(\theta,t)D(\theta)d\theta$ [16], where $\rho(\theta,t)$ is the time dependent angular distribution after integration over the azimuthal angle and including the $\sin\theta$ weight factor. With a sufficiently strong rotational alignment in the molecular sample, the phase of the total harmonic emission can be reversed compared with that from a randomly oriented distribution.

Although double-focus interferometric geometry has been used before to measure the coherence of HHG and the atomic dipole phase of HHG [17, 18], here we use a new geometry that is both exceedingly stable and simple to implement. Beam path-length stability is especially critical for interferometric measurements due to the short wavelength of the harmonic light. We use two glass plates tilted at slight angles, and partially inserted into the focused laser beam, to split the focus into two elliptical focal spots with a diameter of ~100 μm, as shown in Fig. 1. The duration of this pulse is 25fs, with 550 μJ of energy in each focal spot. The harmonics from these two regions emerge and interfere in the far field. Slight adjustment of the angles of the two plates can change the distance between the two foci, and the relative angle between them can change the time delay. A second beam with the same polarization and containing 300 μJ of energy in 120 fs is focused non-collinearly into the molecular gas at the position of one focus, at an intensity of $5 \sim 6 \times 10^{13} W/cm^2$, to create a transient molecular alignment. By blocking the high harmonic generation laser pulse in the unaligned region, we can characterize the harmonic emission from the aligned molecules. A flat-field cylindrical focus grating spectrometer dispersed the harmonic spectrum and imaged the harmonic in one dimension onto an EUV CCD. Stable EUV interference fringes were observed for harmonic orders up to order 33, since the divergence of the HHG emission from the two regions causes a spatial overlap on the CCD in the non-imaging direction. Above order 33, the divergence and intensity of the beams is too small in our geometry to observe fringes.

Fig. 2(a) shows the net harmonic intensity for orders 21-47 as a function of time delay between the alignment and harmonic generation pulses through the ¾ revival. These data exhibit a very different behavior for harmonics above and below the 29th. To see this more clearly, lineouts of different harmonics are shown in Figs. 2 (c) - (e). Below order 29, the intensity follows an inverse of the $<\cos^2\theta>$ ensemble averaged alignment parameter (shown in Fig. 2(b)); i.e. HHG emission is minimum when the molecule is aligned parallel to the laser polarization. In contrast, for harmonics > 29, the yield first goes through a minimum, and then increases for a short time when the molecules are best aligned. This anomalous peak has not been reported before, and is strongly suggestive of a phase shift due to quantum interferences.

To confirm that quantum interferences are the origin of the additional central peak in the HHG from aligned molecules, we directly compare the phase of HHG from aligned and unaligned molecules using the setup shown in Fig. 1. Fig. 3 (a) and (c) plot the observed interference pattern for orders 27 and 33. For order 33, a fringe shift can clearly be observed when the molecules are strongly aligned. No such shift is observed for order 27. The time window for this phase shift exactly matches the duration of the anomalous peak seen in the intensity of order 33 as a function of alignment (Fig. 2 (d)). To determine the magnitude of the phase shift between harmonic emission from the aligned and randomly-oriented molecules, we integrate the interferences fringes between -100fs and 100fs. We then compare this with the integrated fringes outside this temporal window, where there is no strong alignment. The results are shown in Fig. 3 (b) and (d). We use a sine function added to a slowly varying polynomial background to fit the data. We retrieve a phase difference of $3.4 \pm 0.3$ radians for order 33, and $0.035 \pm 0.5$ radians for order 27.

The inset of Fig. 2 (b) shows the predicted angular distribution at three different times in the ¾ revival, to illustrate the effect of angular averaging. As explained above, the critical angle $\theta_c$ corresponds to that angle where the phase of the harmonic emission changes. At maximum alignment (red curve), most of the distribution is on one side of $\theta_c \approx 34°$ for order 33. At the other two time delays (blue and black curve), the net harmonic signal will have opposite phase since most of that distribution lies on the other side of $\theta_c$. Therefore, a phase shift of $\pi$ is expected in the interference pattern during the revival, in agreement with our data. In contrast, for harmonics < 29 the critical angle is smaller, and therefore the phase of the harmonic emission does not change during the revival. Putting this result into the context, in Ref. [12, 13] the authors convolved the two-center harmonic intensity modulation with the angular distribution *without* accounting for the coherent nature of HHG generation. Our work shows clearly that the harmonic emission adds coherently.

From our measurements of the harmonic intensity and phase, we accurately identify the spectral position of the interference minimum and corresponding phase shift in harmonic emission, which qualitatively agree with the two-center interference model. To further test the validity of this model, we fit the measured harmonic intensities for times around the 3/4 revival using the expression -

$$HHG(t) = \left| \int \rho(\theta,t) A \sin(\pi R \cos\theta / \lambda) d\theta \right|^2 + C \quad (1)$$

In Eqn. 1, the $\sin(\pi R \cos\theta / \lambda)$ term results in quantum interference. *A* is a scaling factor, while *C* accounts for any non-coherent part in the detected signal. Since we do not have very accurate independent measures of the rotational temperature and the pump laser intensity, we can regard these as fit parameters for the entire data set. The best overall fit for all harmonic orders was obtained using a rotational temperature of 105 K, and a pump laser intensity of $5.5 \times 10^{13} W/cm^2$, which is within the estimated range based on our experimental parameters. Least squares fits to Eqn. 1 for three representative harmonics - orders 23, 33, and 39 – are shown in Figs. 2 (c) - (e).

These fits are also in excellent agreement with the two-center model, and also provide compelling evidence that the primary features in the angular modulation of HHG emission from $CO_2$ result from quantum interferences in the recombination process [1].

Since the inter-atomic separation $R$ should be independent of the harmonic order, we can use $R$ as a "ruler" by which to measure the wavelength $\lambda$ of the recolliding electron. Fig. 4 plots the extracted value of $R/\lambda$ (blue squares), together with the theoretically predicted curves for $R$=2.32 Å and two different dispersion relationships: $E_k = nh\omega - \delta I_p$ ($\delta$=0 or 1). Our experimental data show that the ratio $R/\lambda$ does not follow a relationship corresponding to a fixed value of $\delta$, but rather that $\delta$ is near 0 for low order harmonics, and increases for higher orders. The effect of the Coulomb potential on the continuum electron is the most probable reason for this behavior. The molecular potential can further accelerate the returning electron and distort the wave front from that of a plane wave. To a first approximation, we can consider the Coulomb effect as a correction to the kinetic energy of the returning electron plane wave. The equation $E_k = nh\omega - I_p$ gives asymptotic kinetic energy of the electron, corresponding to some momentum vector $k$. Further acceleration of the electron inside the molecular potential will give an additional momentum shift $\Delta k$ and a phase shift $\Delta k \cdot R$. Our data show that this phase shift is small for high order harmonic and larger for lower orders. For harmonic orders below 25, our extracted value of $R/\lambda$ is even larger than the calculated value using a dispersion relationship $E_k = nh\omega$, which most likely is due to a breakdown of the plane wave approximation for low order harmonics. Other possible explanations include multi-electron effects, or field distortion of the molecular HOMO which might stretch the effective distance between the centers of electron density. Two dashed curves calculated using a slightly longer $R$ (2.45 Å) are also shown in Fig. 4 for comparison. The effort to develop a complete theory of HHG in molecules that includes multi-electron and Coulomb effects will be helpful to fully understand our data [19, 20].

In summary, we present an accurate and unambiguous measurement of the phase shift of high-order harmonic emission in molecules for the first time. The data also allow for a detailed measurement of the dispersion relation between the wavelength of the recolliding electron and the harmonic order, and will be very useful to benchmark complete theories of harmonic emission from molecules. Future studies can extend this technique to molecules such as $N_2$, that cannot be described as a pure symmetric combination of atomic orbitals [21], where the phase of the EUV emission might depend on the molecular orientation in a more complex way.

The authors gratefully acknowledge support from the US Department of Energy Office of Basic Energy Sciences, and from the National Science Foundation.

# References

[1]     J. Itatani *et al.*, Nature (London) **432**, 867 (2004).

[2]     J. Itatani *et al.*, Phys. Rev. Lett. **94**, 123902 (2005).

[3]     T. Kanai, S. Minemoto, and H. Sakai, Nature (London) **435**, 470 (2005).

[4]     N. L. Wagner *et al.*, Proc. Natl. Acad. Sci. USA **103**, 13279 (2006).

[5]     R. Torres *et al.*, Phys. Rev. Lett. **98**, 203007 (2007).

[6]     J. Ortigoso *et al.*, J. Chem. Phys. **110**, 3870 (1999).

[7]     H. Stapelfeldt, and T. Seideman, Rev. Mod. Phys. **75**, 543 (2003).

[8]     M. Lewenstein *et al.*, Phys. Rev. A **49**, 2117 (1994).

[9]     H. D. Cohen, and U. Fano, Phys. Rev. **150**, 30 (1966).

[10]    M. Lein *et al.*, Phys. Rev. Lett. **88**, 183903 (2002).

[11]    M. Lein *et al.*, Phys. Rev. A **66**, 023805 (2002).

[12]    C. Vozzi *et al.*, Phys. Rev. Lett. **95**, 153902 (2005).

[13]    A. T. Le, X. M. Tong, and C. D. Lin, Phys. Rev. A **73**, 041402 (2006).

[14]    H. Wabnitz *et al.*, Eur. Phys. J. D **40**, 305 (2006).

[15]    N. L. Wagner *et al.*, Submitted (2007).

[16]    S. Ramakrishna, and T. Seideman, Phys. Rev. Lett. **99**, 113901 (2007).

[17]    M. Bellini *et al.*, Phys. Rev. Lett. **81**, 297 (1998).

[18]    C. Corsi *et al.*, Phys. Rev. Lett. **97**, 023901 (2006).

[19]    S. Patchkovskii *et al.*, J. Chem. Phys. **126**, 114306 (2007).

[20]    O. Smirnova *et al.*, J. Phys. B **40**, F197 (2007).

[21]    B. Zimmermann, M. Lein, and J. M. Rost, Phys. Rev. A **71**, 033401 (2005).

**Figure Captions**

**Figure 1:** Setup for directly measuring the intensity and phase of high harmonic emission from molecules. The two foci are 240 μm and 110 μm away from the gas jet exit respectively. The diameter of the jet orifice is 150 μm and the backing pressure is 700 Torr. HHG from aligned and randomly oriented molecules from two different regions interfere in the far field.

**Figure 2:** (a) Experimentally measured intensity (from the aligned molecular sample only) for harmonic orders 21 - 47 as a function of time-delay between the aligning laser pulse and harmonic-generating pulse within the ¾ revival. Time zero is shifted to the center of this revival for convenience. (b) Predicted $<\cos^2 \theta>$ alignment assuming a rotational temperature of 105 K. The inset plots the angular distribution at three selected times within the revival, color coded to indicate the corresponding time on the $<\cos^2 \theta>$ graph. Here, 0° corresponds to alignment along the laser polarization, while 90° corresponds to "anti-alignment." (i.e. perpendicular to the laser polarization) The critical angle of 34° for harmonic order 33 is also labeled. (c) – (e) Lineout of harmonic orders 23, 27, 33, 37, 39, and 41 that exhibit different sub-structure in the harmonic emission at optimal alignment. The dashed line shows a least-square fit to Eqn. 1 for harmonic orders 23, 33 and 39.

**Figure 3:** (a), (c) Interference pattern as a function of time within the ¾ revival for harmonic orders 27 (a) and 33 (c). (b) Intensity-scaled integrated fringes for the $27^{th}$ harmonic in the -100fs to 100fs interval (red squares), along with least square fit (red solid line). Integrated fringes outside this temporal window are also shown (black circles) as well as a least square fit (black solid line). (d) Same as (b), but for the $33^{rd}$ harmonic.

**Figure 4:** Extracted value of $R/\lambda$ versus harmonic order from the fits shown in Figs. 2 (c) - (e) (blue squares with error bar). The calculated value of $R/\lambda$ is also shown for two different dispersion relationships and internuclear separations: $E_k = nh\omega - I_p$ and $R$=2.32 Å (black solid line), R=2.45 Å (black dashed line); $E_k = nh\omega$ and $R$=2.32 Å (red solid line), R=2.45 Å (red dashed line).

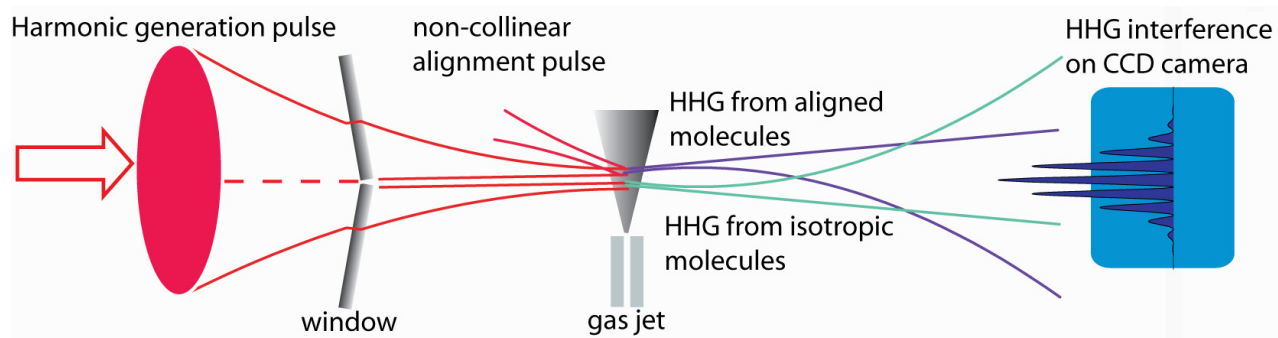

**Figure 1**

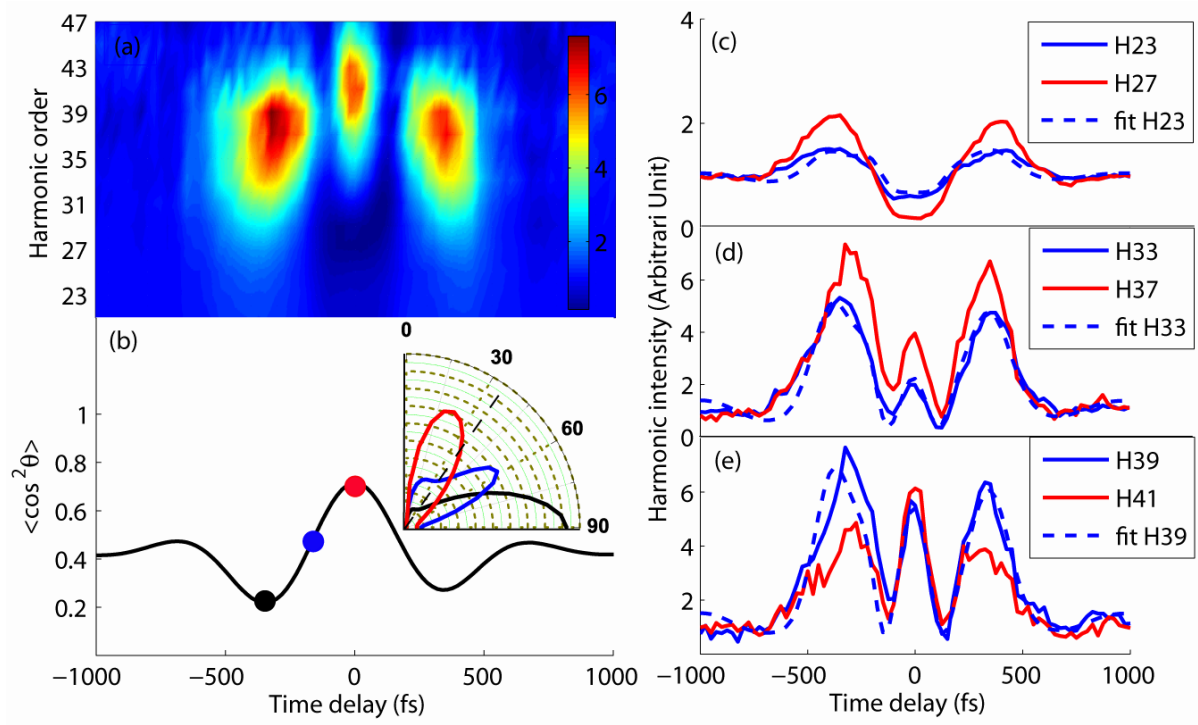

**Figure 2**

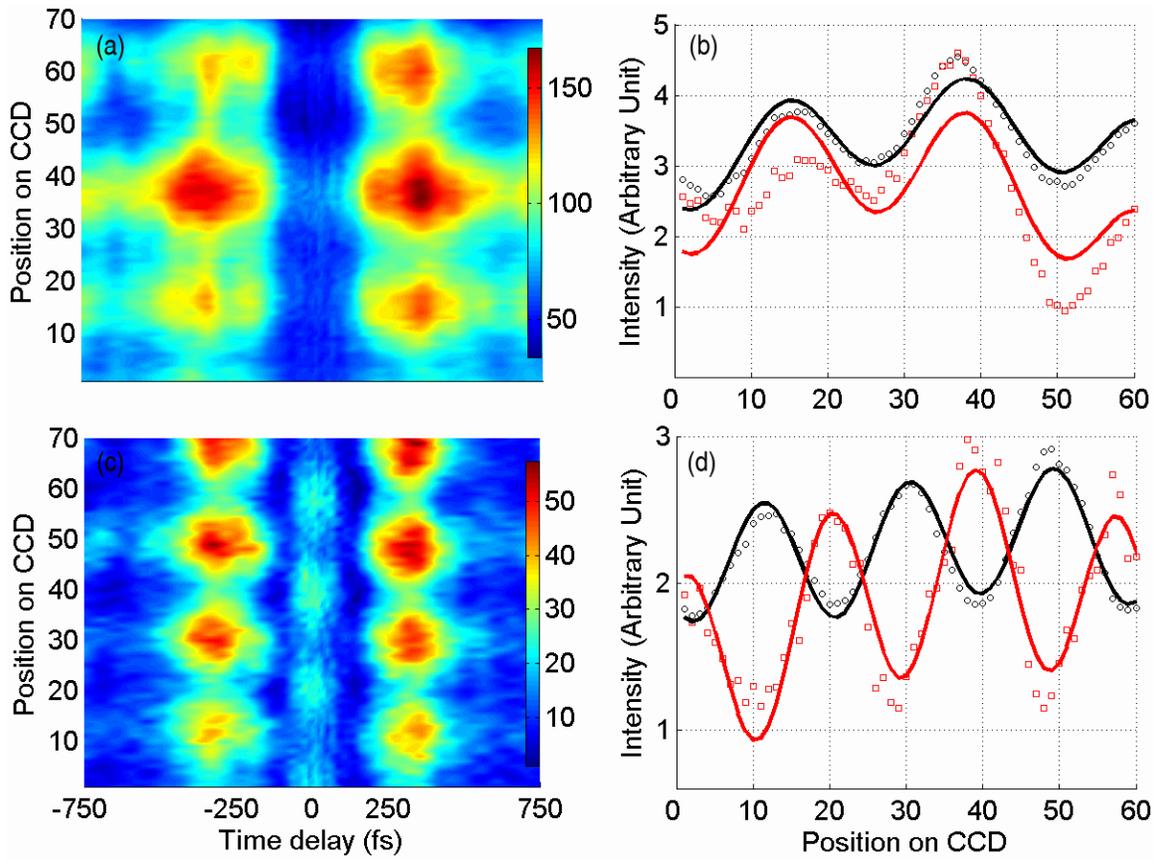

**Figure 3**

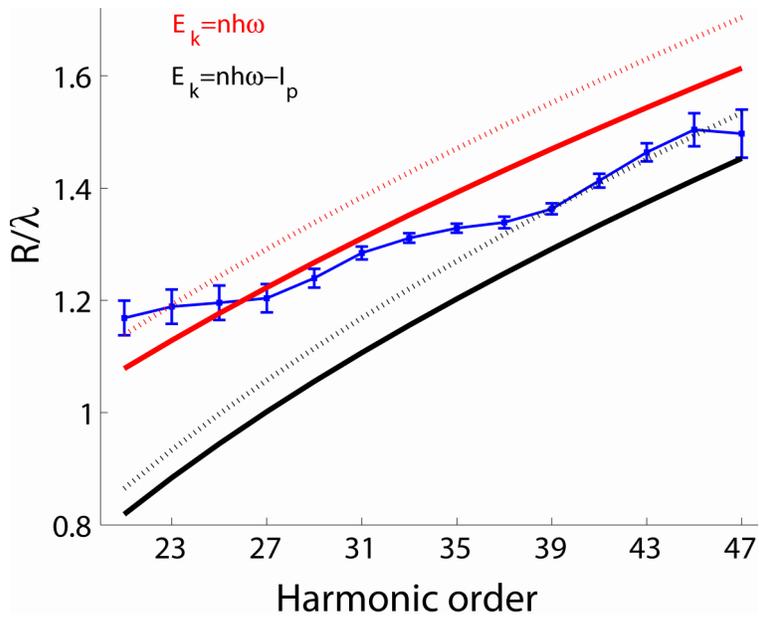

**Figure 4**